\documentclass[runningheads]{llncs}
\usepackage{graphicx}
\usepackage[pdftex,dvipsnames]{xcolor} 
\usepackage{xargs}
\usepackage{amssymb}
\usepackage{hyperref}
\usepackage{wrapfig}
\usepackage[colorinlistoftodos,prependcaption,textsize=tiny]{todonotes}

\usepackage[caption=false]{subfig}
\usepackage{pdfpages}

\usepackage{etoolbox}
\newtoggle{arxiv}
\newcommand{\ifarxivelse}[2]{\iftoggle{arxiv}{#1}{#2}}
\toggletrue{arxiv} 

\usepackage[firstpage]{draftwatermark}
\SetWatermarkText{\hspace*{5in}\raisebox{8in}{\includegraphics[scale=0.1]{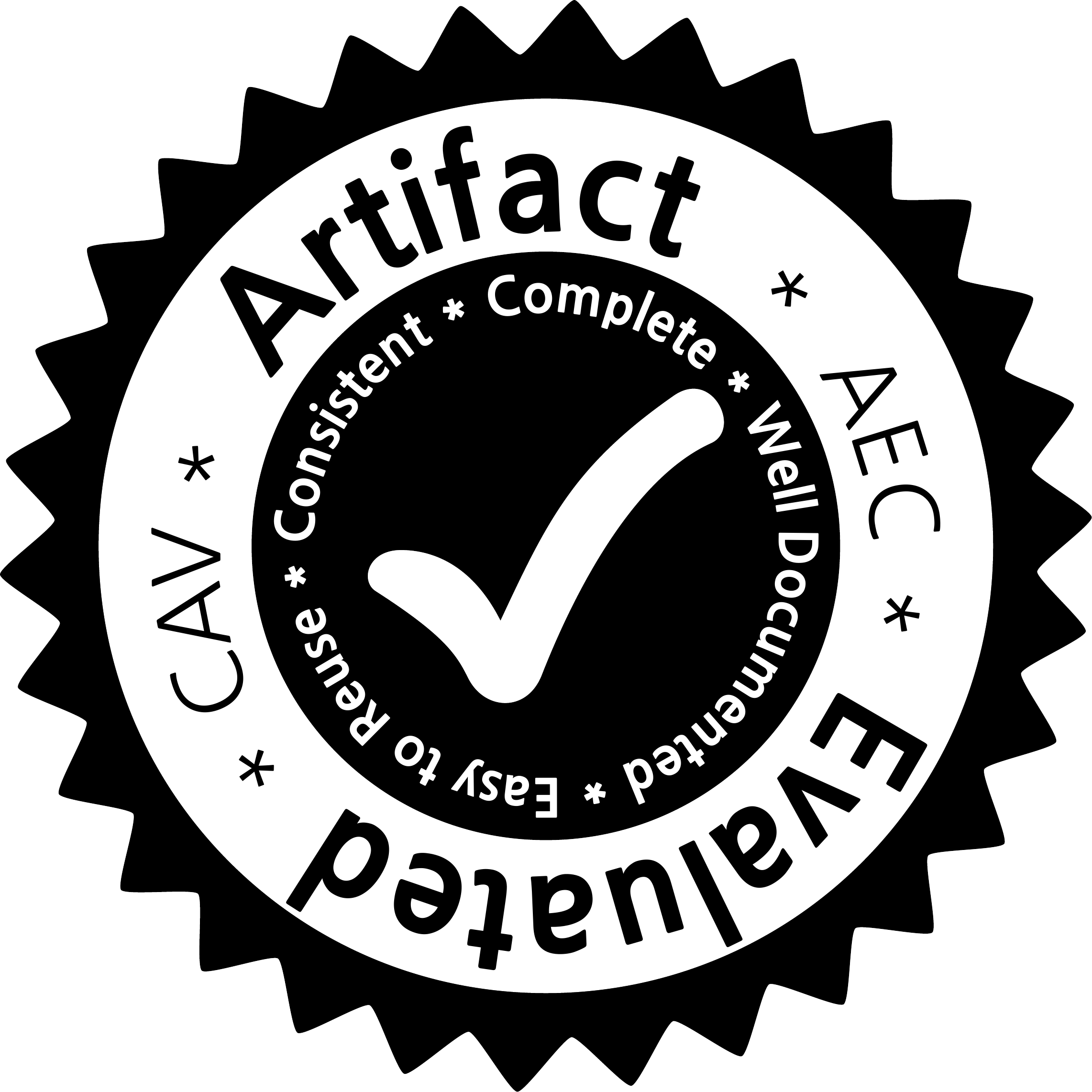}}}
\SetWatermarkAngle{0}

\newcommand{\subsubsectionx}[1]{\vspace*{-0.0em}\framebox[1.1\width]{\bf #1}~\vspace*{-0.5em}}

\newcommand{\myspace}{\vspace*{-0.5em}}

\newcommand{\mypar}[1]{\vspace*{-0.5em}\paragraph{#1}}

\begin{document}

\title{Automata Tutor v3\thanks{
		We thank Emil Ratko-Dehnert from ProLehre TUM for the professional help with the student survey; Tobias Nipkow and his team for allowing us to conduct the user survey in his class; Christian Backs, Vadim Goryainov, Sebastian Mair and Jan Wagener for the exercises they added as part of their Bachelor's theses; Julia Eisentraut and Salomon Sickert-Zehnter for their help in developing this project; the TUM fund ``Verbesserung der Lehrmittelsituation'' and the CAV community for caring about good teaching. 
		Loris D'Antoni was supported, in part, by NSF under grants CNS-1763871, CCF-1750965, CCF-1744614, and CCF-1704117; and by the UW-Madison OVRGE with funding from WARF.
}}

%
%
\author{
	Loris D'Antoni\inst{1}\and
	Martin Helfrich\inst{2}\and
	Jan Kretinsky\inst{2}\and
	Emanuel Ramneantu\inst{2}\and
	Maximilian Weininger\inst{2}
}
\authorrunning{L. D'Antoni et al.}
%
\institute{University of Wisconsin-Madison \\
	\email{loris@cs.wisc.edu }\\
	 \and
Technical University of Munich\\
\email{\{martin.helfrich,jan.kretinsky,emanuel.ramneantu,maxi.weininger\}@tum.de}\\
}
\maketitle              

\myspace\myspace

\begin{abstract}
Computer science class enrollments have rapidly risen in the past decade.
With current class sizes, standard approaches to grading and providing personalized feedback are no longer possible and new techniques become both feasible and necessary.
In this paper, we present the third version of Automata Tutor, a tool 
for helping teachers and students in large courses on 
automata and formal languages.
The second version of Automata Tutor supported automatic grading and feedback for finite-automata constructions and
has already been used by thousands of users in dozens of countries.
This new version of Automata Tutor supports automated grading and feedback generation 
for a greatly extended variety of new
problems, including problems that
ask students to create regular expressions, context-free grammars, pushdown automata and Turing machines corresponding
to a given description, and problems about converting between equivalent models - e.g., from regular expressions to nondeterministic finite automata.
Moreover, for several problems, this new version also enables teachers and students to automatically generate new problem
instances.
We also present the results of a survey run on a class of 950 students, which shows
very positive results about the usability and usefulness of the tool.
\end{abstract}

\myspace\myspace
\myspace\myspace\myspace

\section{Introduction}

\myspace

Computer science (CS) class enrollments have been rapidly rising, e.g., CS enrollment roughly triples per decade at Berkeley and Stanford
\cite{Pat13} or TU Munich.
Both online and offline courses and degrees are being created to educate students and professionals in computer science
and these courses may soon have thousands of students attending a lecture, or tens of thousands following a Massive Online Open Course (MOOC). At these scales, standard approaches to grading and providing personalized feedback are no longer possible and new techniques become both feasible and necessary.
Current approaches for handling this growing student volume include reducing the complexity of assignments or relying on imprecise feedback and grading mechanisms. Simpler assessment mechanisms, e.g., multiple-choice questions, are easier to grade automatically but lack realism~\cite{hpl2000}.
Designing better techniques for automated grading and feedback generation is therefore
a necessity.

Recent advances in formal methods, including program synthesis and verification, can help teachers and students in verifiably correct ways that statistical or rule-based techniques cannot. 
For example, formal methods have been used to identify student errors and provide feedback for
problems related to
introductory Python programming assignments~\cite{SGS13}
 geometry~\cite{Gulwani11,itz13}, algebra~\cite{SinghGR12}, logic~\cite{AhmedGK13}, and automata~\cite{ijcai13,tochi15}. 
 In particular, for this last topic,
the tool Automata Tutor v2~\cite{AToriginal} has already been used 
by more than 9,000 students 
at more than 30 universities in North America, South America, Europe, and Asia.

In this paper, we present Automata Tutor v3, an online\footnote{\url{https://automata.model.in.tum.de}} tool  
that extends Automata Tutor v2 and
uses techniques from program synthesis and decision procedures to improve the quality and effectiveness of teaching
courses on 
automata and formal languages. Besides
being part of the standard CS curriculum, the concepts taught in these
courses are rich in structure and applications, e.g., 
in
control theory, text editors, lexical analyzers, or models of software interfaces.
Concrete topics in such curricula include
automata, regular expressions, context-free grammars, and Turing machines.
For problems and assignments related to these topics Automata Tutor v3 can automatically:
(1) Detect whether the student's solution is  correct.
(2) Detect different types of student's mistakes and translate them
into explanatory feedback.
(3) If possible, generate new problems together with the corresponding solutions for teachers to use in class.

Automata Tutor v3 greatly expands its predecessor Automata Tutor v2, which  only provides
ways to pose and solve problems for deterministic and nondeterministic finite automata constructions. 
This paper describes the new components introduced by Automata Tutor v3 and how
this new version improves on its previous one.  
The key advantages to its competitors are the breadth, automatic generation and grading of exercises, infrastructure allowing for use in large courses and a useful feedback to the students, compared to text-based interfaces used by Autotool \cite{autotool}, rudimentary feedback in JFLAP \cite{DBLP:conf/ic3/ShekharAABK14} and none in Gradience \cite{Gradience}.

Since Automata Tutor has already been well received by teachers around the world, we believe that
the readers from the CAV community will find great value in knowing about this new and fundamentally richer version of the tool and how it can 
extensively help with teaching the automata and formal languages courses, a task we know many of the attendees have to
face on a yearly basis.
Our contributions are the following:
\begin{itemize}
 \item \textbf{Twelve new types of problems} (added to the four problems from the previous version) that can be created by teachers and for which the \emph{tool
 can assign grades together with feedback} to student attempts. 
 While the previous version of Automata Tutor could only support problems involving finite automata
 constructions, Automata Tutor v3 now supports problems for proving language non-regularity using the pumping lemma, building regular expressions, context free grammars, pushdown automata and Turing machines, and conversions between such models.
 \item \textbf{Automatic problem generation} for five types of problems, with the code modularity allowing to add it for all the others. This feature allows teachers to effortlessly create new assignments, or students to practice by themselves with potentially infinitely many exercises.
 \item A new and \textbf{improved user interface} that allows teachers and students to navigate the increased number of problem types and assignments. Furthermore, each problem type comes with an intuitive user interface (e.g., for drawing pushdown automata).
 \item An improved \textbf{infrastructure} for the use in large courses, in particular, incorporating  login systems (e.g. \emph{LDAP} or \emph{OAuth}), getting a certified mapping from users to students and enabling teachers to grade homework or exams. 
 \item A \textbf{user study} run on a class of 950 students to assess the effectiveness and usability of Automata Tutor v3. 
In our survey, students report to have \emph{learned quickly}, \emph{felt confident}, and \emph{enjoyed} using Automata Tutor v3, and found it \emph{easy to use}. Most importantly, students found the feedback given by the tool to be \emph{useful} and claimed
they \emph{understood more} after using the tool and felt \emph{better prepared} for an upcoming exam.
In our personal experience, the tool saves us dozens of thousands of corrections in each single course.
\end{itemize}


\myspace\myspace\myspace

\section{Automata Tutor in a nutshell}

\myspace

Automata Tutor is an online education tool created to support courses teaching basic concepts in 
automata and formal languages~\cite{AToriginal}.
In this section, we describe how Automata Tutor helps teachers run large courses
and students learn efficiently in such courses.

\begin{figure}[htbp]
\centering
	\includegraphics[width=0.7\linewidth]{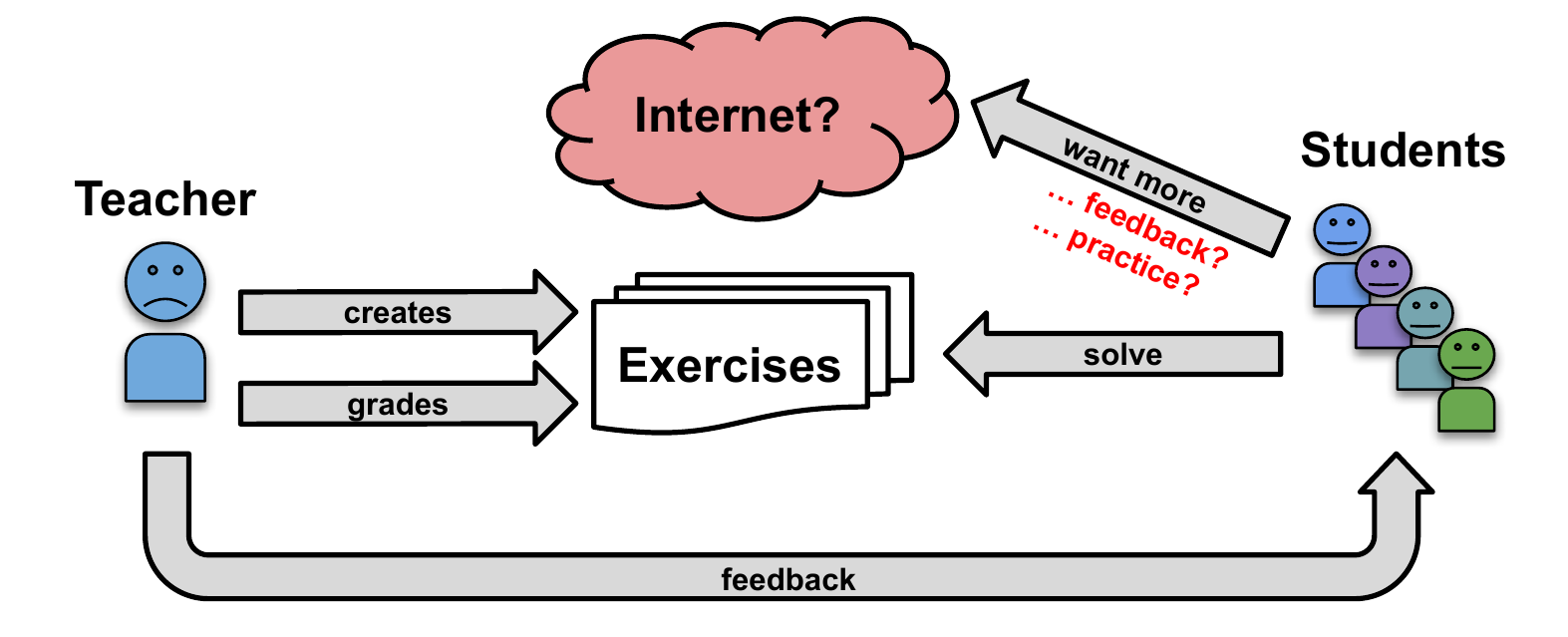}
	\caption{Common structure of practical sessions for CS classes.}
	\label{fig:before_AT}
\end{figure}

\mypar{Learning without Automata Tutor}\quad
Figure \ref{fig:before_AT} schematically shows a student-teacher interaction in a course taught without an online tutoring
system.
The teacher creates exercises, grades them manually, and (sometimes) manually provides
personalized feedback to the students. 
This type of interaction has many limitations:
(1) it is asynchronous (i.e., the student has to wait a long time for what is often little feedback) and does not scale to large classrooms, posing strenuous amount of work on teachers,
(2) it does not guarantee consistency in the assigned grades and feedback, and
(3) it does not allow students to revise their solutions upon receiving feedback as the teachers often release
a solution to all students as part of the feedback and do not grade new submissions.

Another drawback of this interaction is the  limited number of problems students can practice on. 
Because teachers do not have the resources to create many practice problems  and provide feedback for them,
students are often forced to search the Internet for old exams and practice sheets or even exercises from other universities. Due to the lack of feedback, this chaotic search for practice problems often ends up confusing the students
rather than helping them. 
\begin{figure}[htbp]
\centering
	\includegraphics[width=0.8\linewidth]{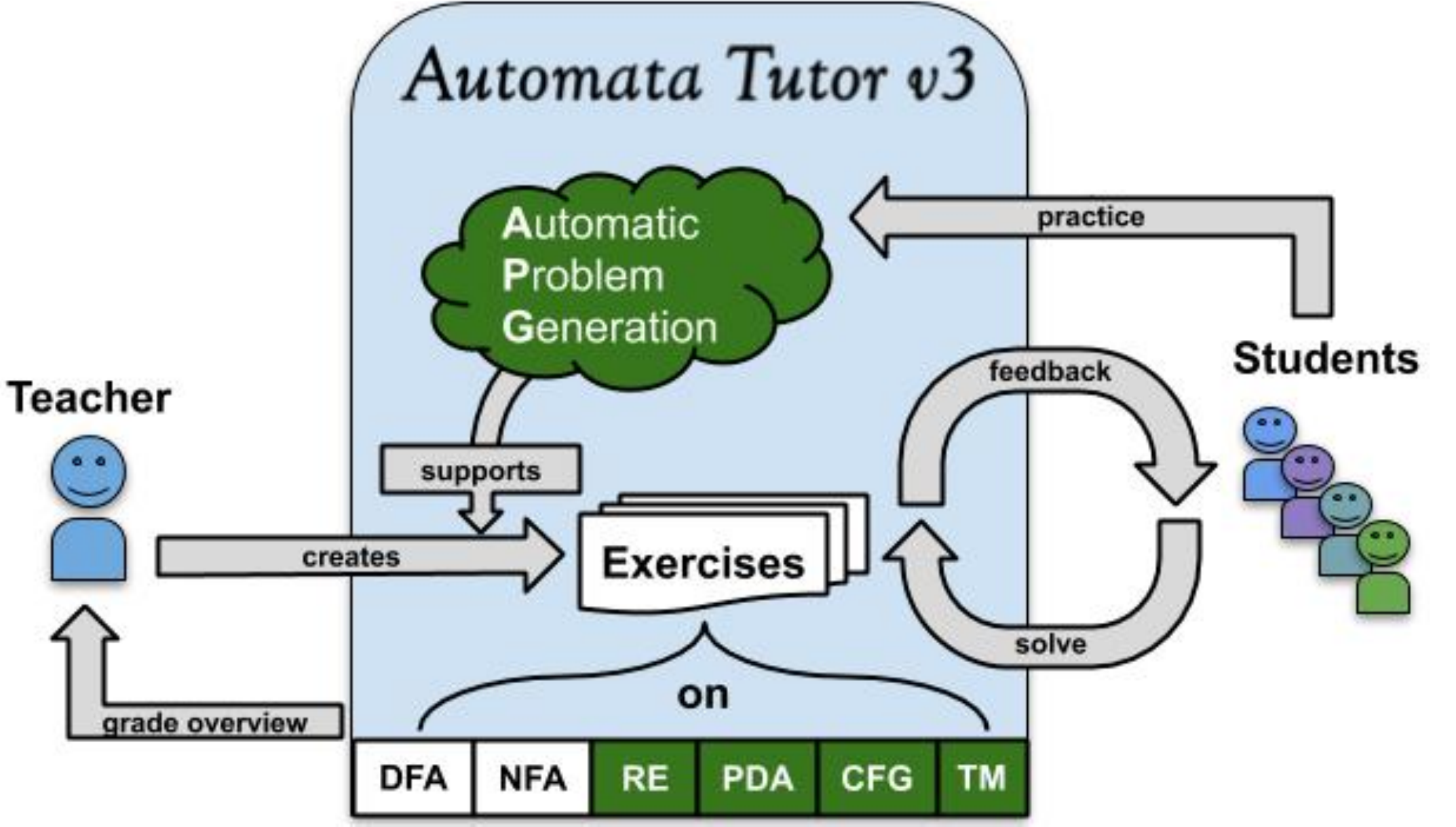}
	\caption{Overview of Automata Tutor v3 (our contributions in green). 
		The teacher creates exercises on various topics. 
		The students solve the exercises in a feedback cycle: 
		After each attempt they are automatically graded and get personalized feedback. 
		The teacher has access to the grade overview. 
		For additional practice, students can generate an unlimited number of new exercises using the automatic problem generation.}
	\label{fig:AT_overview}
\end{figure}


\begin{figure}[!ht]
	\centering
	\includegraphics[width = 1.0\textwidth]{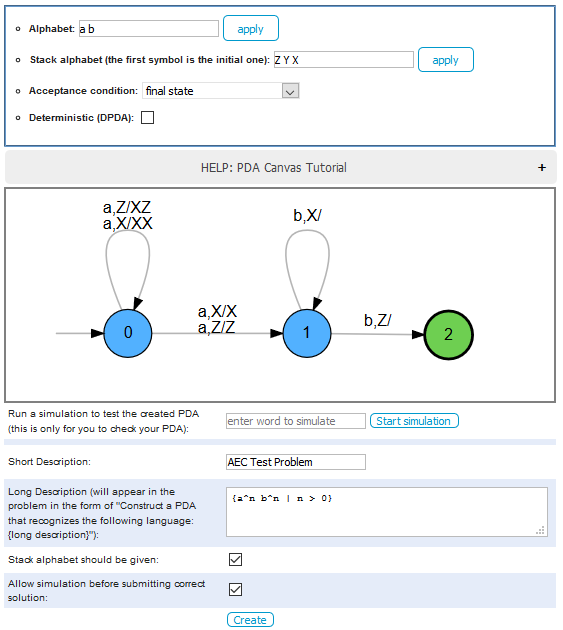}
	\caption{Creating a new problem of type ``PDA Construction''.}
	\label{fig:scCreate}
\end{figure}

\begin{figure}[ht] 
	\centering
	\includegraphics[width = 0.9\textwidth]{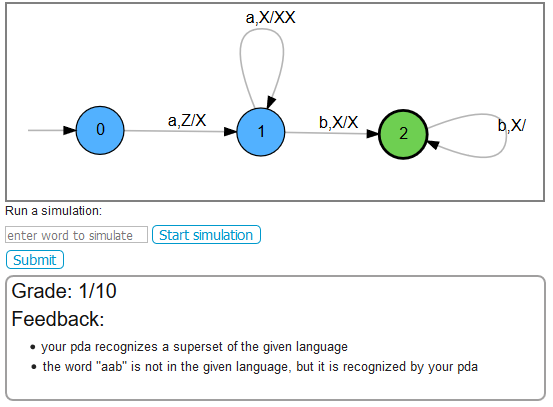}
	\caption{Feedback received when solving the problem created in Figure \ref{fig:scCreate}.}
	\label{fig:scSolve}
\end{figure}

\mypar{Learning with Automata Tutor}\quad
Figure \ref{fig:AT_overview} shows the improved interaction offered by Automata Tutor v3.
Here, a teacher creates the problem instances with the help of the tool.
The problems are then posed to the students and, \emph{no matter how large a class is},
Automata Tutor automatically grades the solution attempts of students right when they are submitted and immediately gives detailed and personalized feedback for each submission. 
If required, e.g. for a graded homework, it is possible to restrict the number of attempts.
Using this feedback, the students can immediately try the problem again  and learn from their mistakes. 
As shown in a large user study run on the first version of Automata Tutor~\cite{tochi15},
this fast feedback cycle is encouraging for students and results in students spontaneously exploring
more practice problems and engaging with the course material.
Additional practice is supported by the automatic problem generation, with the same level of detailed and personalized feedback as before without increasing the workload of the teacher. 
Furthermore,  automatic problem generation can assist the teacher in creating new exercises.
Finally, whenever necessary, the teacher can download an overview of all the grades.

\mypar{Improved user interface}\quad
Automata Tutor is an online tool which runs in the most used browsers.
A new collapsible navigation bar groups problems by topic, facilitating quick access to exercises and displaying the structure of the course (see Figure \ifarxivelse{\ref{fig:scNav} in Appendix B}{6 in \cite[Appendix B]{techreport}}).
To create a new exercise, a teacher clicks the ``+''-button and is presented the view of Figure \ref{fig:scCreate}.
In this case, the drawing canvas allows to easily specify the sample solution pushdown automaton.
Similarly, when students solve this exercise, they draw their solution attempt also on the canvas. 
After submitting, they receive their personalized feedback and grade (see example in Figure \ref{fig:scSolve}).
For the automatic problem generation, a dropdown menu to select the problem type and a slider to select the difficulty is displayed together with the list of all problems the user has generated so far (see the screenshot in Figure \ifarxivelse{\ref{fig:scAuto} in Appendix B}{7 in \cite[Appendix B]{techreport}}).


\myspace\myspace

\section{Design}

\myspace

\subsection{University and course management}

\myspace

While Automata Tutor can be used for independent online practice, one of the main advantages is its infrastructure for large university courses.
To this end, it is organized in \emph{courses}. 
A course is created and supervised by one or more teachers. Together, they can create, test and edit exercises. 
The students cannot immediately see the problems, but only after the teachers have decided to pose them.
This involves setting the maximum number of points, the number of allowed attempts as well as the start and end date.

To use Automata Tutor, students must have an account. One can either register by email or, in case the university supports it, login with an external login service like \emph{LDAP} or \emph{Oauth}. When using the login service of their university, teachers get a certified mapping from users to students and enabling teachers to use Automata Tutor v3 for grading homework or exams. 

Students can enroll in a course using a password. Enrolled students see all posed problems and can solve them (using the 
allowed number of attempts). 
The final grade can be accessed by the teachers in the grade overview.

\myspace

\subsection{New problem types}

\myspace

In this section, we list the problem types newly added to Automata Tutor v3.
They are all part of the course \cite{hopcroftUllman} and
a detailed description of each problem can be found in \ifarxivelse{Appendix \ref{app:problems}}{\cite[Appendix A]{techreport}}, including the basic theoretical concept, how a student can solve such a problem, what a teacher has to provide to create a problem, the idea of the grading algorithm, and what feedback the tool gives.

\begin{description}
	\item \textit{RE/CFG/PDA Words:} Finding words in or not in the language of a regular expression, context free grammar or pushdown automaton.
	\item \textit{RE/CFG/PDA Construction:} Given a description of a language, construct a regular expression, context free grammar or pushdown automaton.
	\item \textit{RE to NFA:}  Given a regular expression, construct a nondeterministic-finite automaton.
	\item \textit{Myhill-Nerode Equivalence Classes:} There are two subtypes: either, given a regular expression and two words, find out whether they are equivalent w.r.t. the language, or, given a regular expression and a word, find further words in the same equivalence class.
	\item \textit{Pumping-Lemma Game:} Given a language, the student has to guess whether it is regular or not and then plays the game as one of the quantifiers.
	\item \textit{Find Derivation:} Given a context free grammar and a word, the student has to specify a derivation of that word.
	\item \textit{CNF:} Given a context free grammar, the student has to transform it into Chomsky Normal Form.
	\item \textit{CYK:} Given a context free grammar in CNF and a word, the student has to decide whether the word is in the language of the grammar by using the Cocke–Younger–Kasami  algorithm.
	\item \textit{While to TM:} Given a while-program (a Turing-complete programming language with very restricted syntax), construct a (multi-tape) Turing machine with the same input-output behaviour.
\end{description}

\subsection{Automatic problem generation}\label{sec:autogen}
\newcommand{\qual}{\mathit{qual}}
\newcommand{\diff}{\mathit{diff}}

\myspace

Automatic Problem Generation (APG) allows one to generate new exercises of a requested \emph{difficulty} level and problem type. 
This allows students to practice independently and supports teachers when creating new exercises. 
While APG is currently implemented for
four CFG problem types and for the problem type ``While to TM'', it can be easily extended to other problem types by providing the following components:
\begin{itemize}
	\item \textbf{Procedure for generating exercises at random} either from given basic building blocks or from scratch.
	\item \textbf{A ``quality'' metric} $\qual(E)$ for assessing the quality of the generated exercise $E$, ranging from trivial or infeasible to realistic.
	\item \textbf{A ``difficulty'' metric} $\diff(E)$ for assessing the difficulty of $E$. 
\end{itemize}
Given these components, Automata Tutor generates a new problem with a given minimum difficulty $d_{\min}$ and maximum difficulty $d_{\max}$ as follows. 
Firstly, 100 random exercises are generated.
Secondly, Automata Tutor chooses exercises $E$ with the best quality such that $d_{\min} \leq \diff(E) \leq d_{\max}$.

Concretely, for the CFG problem types, CFGs with random productions are generated and sanitized.
Resulting CFGs that do not accept any words or have too few productions are excluded using the quality metric. 
The difficulty metric always depends on the number of productions; additionally, depending on the exact problem type, further criteria are taken into account.

For the problem type ``While to TM'' we use an approach similar to the one suggested in existing tools for automatic problem generation~\cite{autoGenerate,autoGenerate2}:
We handcrafted several \emph{base programs} which are of different difficulty level. 
In the generation process, the syntax tree of such a base program is abstracted and certain modifying operations are executed; these change the program without affecting the difficulty too much. E.g. we choose different variables, switch the order of if-else branches or change arithmetic operators.
Then several programs are generated and those of bad quality are filtered out.
A program is of bad quality if its language is trivially small or if it contains infinite loops; since detecting these properties is undecidable, we employ heuristics such as checking that the loops terminate for all inputs up to a certain size with a certain timeout.

\myspace\myspace

\section{Implementation and scalability} \label{sec:userStudy}

\myspace

Automata Tutor v3 is open source and it consists of a frontend, a backend, and a database. 
It also provides a developer's manual for creating new exercises.

The frontend, written in {\tt scala}, renders the webpage.
The drawing canvases for the different automata and the Turing machines rely on javascript.
The frontend and backend communicate using XML objects.

The backend, written in C\#, contains methods to unpack the xml of the frontend to compute the grade and feedback for solutions. It is also used to check the syntax of exercises and for the automatic problem generation. 
It relies on AutomataDotNet\footnote{\url{https://github.com/AutomataDotNet/Automata}}, a library that provides efficient algorithms for automata and regular expressions.

The database keeps track of existing users, problems and courses. It uses the H2 Database Engine.

All the new parts of Automata Tutor v3 were developed and tested over the last 3 years at TU Munich, where they were used to support the introductory theoretical computer science course. This local deployment served as
an important test-bed
before publicly deploying the tool online at large scale.
Due to its modular structure, the tool is easily scalable by having multiple frontends and backends together with a load distributor. 
This approach has successfully scaled to 950 concurrent student users;
for this, we used 7 virtual machines: 3 hosting frontends, 3 hosting backends (each with 2 cores 2.60GHz Intel(R) Xeon(R) CPU and 4GB RAM), and 1 for load distribution and the database (with 4 such cores and 8GB RAM).
We will scale the number of machines based on need.

\myspace\myspace

\section{Evaluation and user study}\label{sec:userStudy2}

\myspace


 
\subsubsection{Large-class deployment}~
In the latest iteration of the TU Munich course in 2019, we used Automata Tutor v3 (in the following denoted as AT) in a mandatory homework system for a course with about 950 students; the homework system also included written and programming exercises. 
In total, we posed 79 problems consisting of 18 homework and 61 practice problems.
The teachers saved themselves the effort of correcting 26,535 homework exercises, and the students used AT to get personalized feedback for their work 76,507 times. On average, each student who used AT did so 107 times.

\myspace \myspace 
\subsubsection{Student survey results}~
At the end of the course, we conducted an anonymized survey, based on the System Usability Survey \cite{brooke1996sus}.
14.6\% of the students in the course answered the survey, which is an ordinary rate of return for an online questionnaire, especially given that there was no incentive. 
The students were given statements to judge on a  Likert scale from 1 to 5 (strongly disagree to strongly agree).
We define ``The students agreed with the following statement'' to mean that the average and median scores
were at least 4 and less than 10\% of the students chose a score below 3.
Dually, if the students disagreed with the statement with median and average score that was at most 2 and less than 10\% having a score greater than 3, we say that they ``agreed with the negation of the statement''.
For all statements that do not satisfy either of the criteria, we report mixed answers.
The full survey results can be found in \ifarxivelse{Appendix \ref{app:survey}}{\cite[Appendix C]{techreport}}.

\mypar{Usability}~
Regarding the usability of the tool, the students agreed with the following statements:
\begin{itemize}
	\item I quickly learned to use the AT.
	\item I do \emph{not} need assistance to use the AT.
	\item I feel confident using the AT.
	\item The AT is easy to use.
	\item I enjoy using the AT/the AT is fun to use.
\end{itemize}

However, there were lots of valuable suggestions for improvements, many of which we have implemented 
since then.
Moreover, the survey also revealed space for improvement, in particular for streamlining as documented by the following statements where the answers were more mixed:
\begin{itemize}
	\item The AT is unnecessarily complex.
	\item The canvas for drawing is intuitive.
	\item The use of AT is self-explanatory.
\end{itemize}

\mypar{Usefulness}~
Regarding how useful AT was for learning, the students agreed with the following statements:
\begin{itemize}
	\item I understand more after using the AT.
	\item I prefer using the AT to using pen and paper exercises (12.9\% disagreed, but median and average are 4).
	\item The feedback of the AT was helpful and instructive.
	\item The exercises within the AT are well-designed.
	\item The AT fits in well with the programming tasks and written homework.
	\item The AT did \emph{not} hinder my learning.
	\item I feel better prepared for the exam after using AT.
	\item The feedback of the AT was \emph{not} misleading/confusing.
\end{itemize}
Note that there are no statements with mixed or negative answers regarding the usefulness.
Additionally, as shown in Figure \ref{fig:preferredLearning}, when we asked students about their preferred means of learning,
AT gets the highest approval rate, being preferred to written or programming exercises as well as lectures.

\begin{figure}[t!]
	\includegraphics[width=\linewidth]{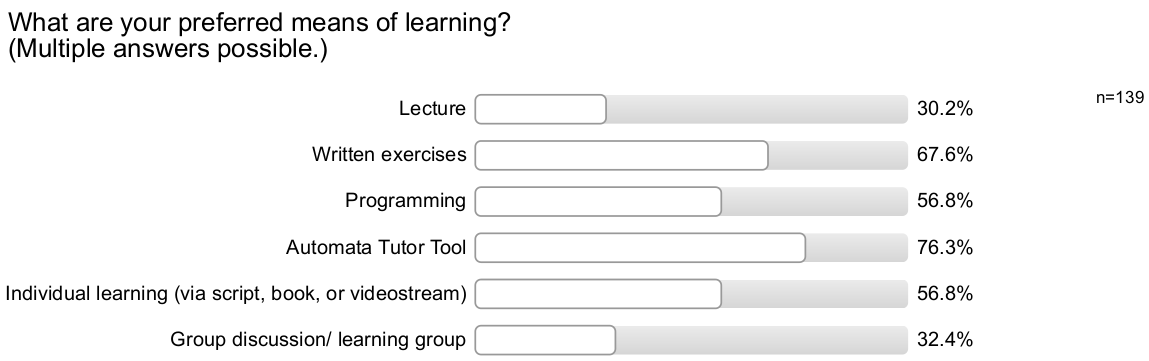}
	\caption{Question from the survey we conducted to evaluate Automata Tutor, showing that the tool is preferred by a majority of students.}
	\label{fig:preferredLearning}
\end{figure}

Overall, this class deployment of Automata Tutor v3 and the accompanying surveys were great successes,
and showed how the tool is of extreme value for both students and teachers, in particular for such large a course.

\myspace

\section{Conclusion}

\myspace


This paper presents the third version of Automata Tutor, an online tool helping teachers and students in large
automata/computation theory courses.
Automata Tutor v3 now supports automated grading and feedback generation 
for a wide variety of problems and, for some of them, even automatic generation of new problem instances.
Furthermore, it is easy to extend and we invite the community to contribute by implementing further exercises.
%
Finally, our experience shows that Automata Tutor v3 improves the economical aspects of teaching greatly as it scales effortlessly with the number of students.

Earlier versions of Automata Tutor have already been adopted by thousands of students at dozens of schools and we hope this paper allows Automata Tutor~v3 to help even more students and teachers around the world.

\bibliographystyle{abbrv}
\bibliography{ref}

\ifarxivelse{
\appendix

\section{Detailed description of the new problem types}\label{app:problems}

In this section, we describe functionality for all the problems that have been newly added to Automata
Tutor v3. We will use the following acronyms: RE: Regular expression, CFG:
Context-free grammar, PDA: Pushdown automaton, NFA: Nondeterministic finite automaton, CNF: Chomsky Normal Form, CYK: Cocke-Younger-Kasami al-
gorithm, TM: Turing machine.
For every type of problem that we added, we will describe the basic theoretical
concept, how a student can solve such a problem, what a teacher has to provide
to create a new problem, the idea of the grading algorithm and the kind of
feedback Automata Tutor v3 gives. 
In the first two subsections, we aggregate
several problem types, as the idea is very similar.

\subsubsectionx{RE/CFG/PDA Words}

\begin{description}
	\item[Content] This type of problem transfers a basic understanding of formal languages and the concept of RE, CFG or PDA~\cite[Chapters 3, 5, 6]{hopcroftUllman}. 
	\item[Solving] The student is given a RE/CFG/PDA and has to provide words that are in the language of the RE/CFG/PDA and words that are not.
	\item[Creating] The teacher provides a RE/CFG/PDA and decides how many words in and how many words not in the language the student should specify.
	\item[Grading] The student gets points for each unique word that was correct. Words that are given multiple times do not count.
	\item[Feedback] The student is informed about each incorrect word.
\end{description}

\subsubsectionx{RE/CFG/PDA Construction}

\begin{description}
	\item[Content] This problem type conveys an intermediate understanding of formal languages and the concept of RE, CFG or PDA. Note, that the earlier version of Automata Tutor (v2) already contains this problem type for deterministic finite automata, NFA and RE. However, we changed the version for RE significantly.
	\item[Solving] The student is given the description of a language in text form, either as natural language or as formal specification, and has to provide a RE/CFG/PDA that recognizes the language.
	\item[Creating] The teacher has to provide the RE/CFG/PDA and its description. It is up to him to check that they indeed match. Two of the problem types have a way to support the teacher in this:\\
	\textbf{For RE:} The teacher is able to specify equivalent REs. 
	Additionally, this supports the grading algorithm, as described below.\\
	\textbf{For PDA:} The teacher is supported by a simulation for PDA, i.e. the possibility to run the PDA on a word and observe the current state and stack at each point during this run.
	\item[Grading] The student gets more points the closer the attempt was to the sample solution. This depends on the learned concept: \\
	\textbf{For RE:} REs recognizing the correct language always get full points. 
	Incorrect REs are not graded based on the difference in the language they recognize, but according to the Levenshtein edit distance~\cite{levenshtein} to any of the possible REs the teacher provided. For every necessary edit, 20\% of the points are deducted. This is preferable to comparing the languages, because a small careless mistake in the RE can have a large impact on the language. 
	The grading then depends on the teacher providing a few sensible REs.\\
	\textbf{For CFG/PDA:} Equivalence for PDA and CFGs is undecidable. Thus, we perform a limited equivalence check for all words up to a certain length.
	This length is not fixed, but depends on the grammar and the size of the alphabet, even for small word length the check might take minutes.
	To guarantee a fast response, the backend is given one second to check as many words as possible.
	It constructs the sets $A$, words accepted by the correct solution, and $B$, words accepted by the student solution.
	The final grade is calculated as $\frac{|A \cap B|}{|A \cup B|}$ times the maximum number of points. 
	This suffices, as the exercises are created and solved by hand, so it is unlikely that a differentiating word is very long.
	Still, even when the full grade is achieved, Automata Tutor does not claim that the languages are equal, but only reports that the student solution passed all tests, giving the amount of words tried.
	\item[Feedback] The student gets counterexamples for the equivalence, i.e. a word that is part of one of the languages but not the other. 
	Additionally, for PDA, the teacher can allow the students to also simulate words in their PDA.
\end{description}



\subsubsectionx{RE to NFA}

\begin{description}
	\item[Content] This problem type teaches the transformation from RE to $\epsilon$-NFA, cementing the understanding of both concepts. 
	It does not only allow to directly specify the solution $\epsilon$-NFA, but also gives the possibility to follow the step-by-step algorithm \cite[Chapter 3.2.3]{hopcroftUllman}, using ``block states'' to resolve one RE constructor at a time.
	\item[Solving] The student is given a RE and has to provide an $\epsilon$-NFA recognizing the same language. 
	To do so, the canvas allows to create normal states and block states.
	\item[Creating] The teacher only has to provide a RE. 
	\item[Grading] 
	The grade is calculated as the number of used block states divided by the number of correct block states times the maximum grade.
	Note that the overall automaton also counts as a block state, so if the student directly specifies an $\epsilon$-NFA without using block states, the grading is binary (0 or full points).
	The rationale for this is, that the exercise incentivizes students to document their path to solution and take small steps; however, we do not want to force students to adhere to a lengthy construction if they already understand the concept.
	\item[Feedback] The feedback informs about block states with incorrect specification; the more the students adhere to the exact algorithm, the more this feedback helps them, as they can more easily identify the position of the error.
	Additionally, whenever a block state with an invalid label is created, e.g. something that is not a subexpression of the goal RE, the student is immediately notified.
\end{description}




\subsubsectionx{Equivalence Classes}

\begin{description}
	\item[Content] This problem type conveys understanding of equivalence classes of a regular language;
	two words $w_1, w_2$ are equivalent with respect to some regular language $L$, if for all possible suffixes $x$ they have the same acceptance behaviour, i.e. $w_1 x \in L \iff w_2 x \in L$~\cite[Chapter 4]{hopcroftUllman}.
	\item[Solving] 
	There are two subtypes of this problem type:
	For the first subtype, the student is given a RE and two words, and has to decide whether they are equivalent or not (with respect to the language of the RE). In both cases a justification is necessary, either the language of suffixes which is accepted after both words or a differentiating suffix.
	For the second subtype, the student is given an RE and a word and has to find further words that are in the same equivalence class (with respect to the language of the RE).
	\item[Creating] The teacher has to provide a RE and decide the subtype of the problem. Depending on the subtype, the teacher has to specify either one or two further words. Additionally, the number of words the student has to provide has to be specified for the second subtype.
	\item[Grading] For the first subtype, the grade depends on the correct assessment of whether the words are equivalent and on the justification.
	For the second subtype, the grading is the same as in problem type ``RE Words''.
	\item[Feedback] For the first subtype, the student is informed whether the assessment was correct. If it was, but the justification is wrong, the feedback either gives information about how the language of the suffixes is different from what the student provided, or whether both words with the suffix appended are in or not in the language.
\end{description}

\subsubsectionx{Pumping-Lemma Game}

\begin{description}
	\item[Content] The point of this problem type is understanding the pumping lemma for regular languages \cite[Chapter 4.1]{hopcroftUllman} via the game where one player instantiates the existential and the other the universal quantifiers. 
	\item[Solving] Given a language, the student states whether the language is regular or not. Afterwards, the student and Automata Tutor play the pumping lemma game against each other: They take turns instantiating the quantified values of the pumping lemma. Depending on the initial choice, the student either initiates the existentially quantified variables (i.e. the pumping lemma number $n$ and the split of the word) or the universally quantifies values (i.e. the word and the number $i$ that describes how often to pump).
	\item[Creating] To specify a language, the teacher is offered the format of arithmetic language. In short, this allows to add exponents to terminals, indicating how often they are repeated, and to give constraints on these exponents; for example in the language 
	$\{ a^i b^j \mid i < j\}$) $i$ and $j$ are the exponents and $i<j$ is a constraint. 
	If the language is not regular, additionally an unpumpable word has to be specified (e.g. $a^n b^{n+1}$). This allows AutomataTutor to win the game if the student chooses ``regular''.
	Note that there are languages that satisfy the pumping lemma, but that are not regular. It is up to the teacher to avoid posing such exercises.
	\item[Grading] The student gets all points for a win and zero points for a loss. Automata Tutor always wins, if the choice of regular/irregular was wrong. If that choice was correct, but further mistakes were made (e.g. the pumping lemma number was chosen too small), Automata Tutor takes advantage and wins the game.
	\item[Feedback] The student sees the choices of Automata Tutor. If Automata Tutor wins, the student can analyze the counterexample. 
\end{description}

\subsubsectionx{Find Derivation}

\begin{description}
	\item[Content] This problem type transfers a basic understanding of (leftmost / rightmost) derivations for CFGs~\cite[Chapter 5.1]{hopcroftUllman}.
	\item[Solving] Given a CFG and an accepting word, the student has to specify any / a leftmost / a rightmost derivation for that word. The derivation has to be given step by step. Each step results from the previous one by replacing a nonterminal according to one of the production rules.
	\item[Creating] The teacher gives a CFG together with a word in its language and decides the type of derivation (i.e. any / leftmost / rightmost).
	\item[Grading] The student gets all points if the derivation was correct and zero points otherwise, i.e. binary grading.
	\item[Feedback] The student is informed about the first incorrect step in the derivation. If the error is due to the replacement of the wrong nonterminal (i.e. not the leftmost / rightmost one), the student gets a corresponding hint.
\end{description}

\subsubsectionx{CNF}

\begin{description}
	\item[Content] This problem type conveys understanding of the Chomsky Normal Form (CNF) for grammars and the transformation algorithm to CNF~\cite[Chapter 7.1]{hopcroftUllman}.
	\item[Solving] Given a CFG, the student needs to find a grammar in CNF that accepts the same language.
	\item[Creating] The teacher gives a CFG.
	\item[Grading] The student gets points according to the equivalence metric of problem type ``CFG Construction'' (i.e. using the limited equivalence test).
	\item[Feedback] First, it is checked if the grammar is in CNF. If the grammar is not in CNF, the attempt is not graded and the student is informed about why the grammar is not in CNF. Otherwise, the student gets counterexamples for the equivalence of two grammars, i.e. words that are only accepted by one of the two grammars.
\end{description}

\subsubsectionx{CYK}

\begin{description}
	\item[Content] Understanding of the CYK algorithm~\cite[Chapter 7.4.4]{hopcroftUllman} that decides if a word is accepted by a CFG.
	\item[Solving] Given a grammar in CNF and a word, the student fills out the CYK-table according to the algorithm.
	\item[Creating] The teacher gives a grammar in CNF and a word.
	\item[Grading] The table is checked row by row starting from the bottom until one row is incorrect. For each correct row the user gets points.
	\item[Feedback] For each wrong cells in the incorrect row, the student gets a hint. Each hint notifies the user that there are nonterminals in the cell that do not belong there and/or that the cell is missing some nonterminal. However, the hints do not contain specific nonterminals.
\end{description}

\subsubsectionx{While to TM}

\begin{description}
	\item[Content] This problem type concerns the conversion of while-programs, a Turing-complete programming language with very restricted syntax, to TMs.
	It helps to understand the connection between the theoretical CS model (TM) and programming, which they already know.
	The exercise looks for equivalence in the input-output-behaviour of while-programs and TMs;
	that means that given input values $x_0, \dots, x_n$ (as value of the variables of the while-program or initial content of the tapes of the TM), after executing the while-program or TM, the output values (variables/tape contents) are the same.
	\item[Solving] The student is given a while-program with $n$ variables. The user interface allows to draw an $n$ tape TM, where every tape corresponds to a variable in the program. 
	The student then has to create a TM that mimics the input-output behaviour of the while-program.	
	\item[Creating] The teacher has to specify a terminating while-program. The teacher should be able to check termination, as the while-programs for which construction a TM is feasible are typically very short and simple.
	\item[Grading] The grading is similar to that of the problem type ``CFG Construction'': Given a certain time, inputs up to a certain length are tested. In order to avoid infinite loops, every input is tested for at most 1000 steps; this value is feasible in terms of running time, because firstly, performing 1000 steps of a Turing machine is not computationally expensive, and secondly, the number of runs that reach this step limit is typically very low.
	\item[Feedback] The feedback informs how many inputs where tested, how many of them were correct. In case there were runs that did not behave as expected, up to five counterexamples are given. This contain the expected and the computed output. It also possible to simulate the TM on the counterexamples.
\end{description}

\section{Additional screenshots}

\begin{figure}
	\centering
	\includegraphics[width=0.33\textwidth]{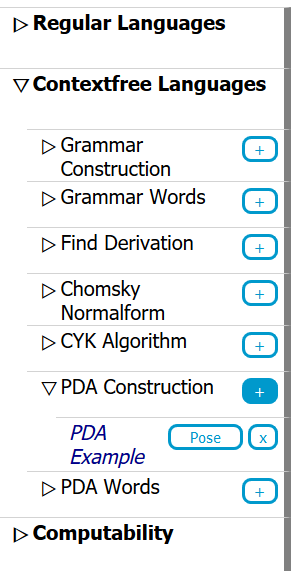}
	\caption{New navigation bar, grouped by topics.}
	\label{fig:scNav}
\end{figure}

\begin{figure}[htbp] 
	\centering
	\includegraphics[width = 1.0\textwidth]{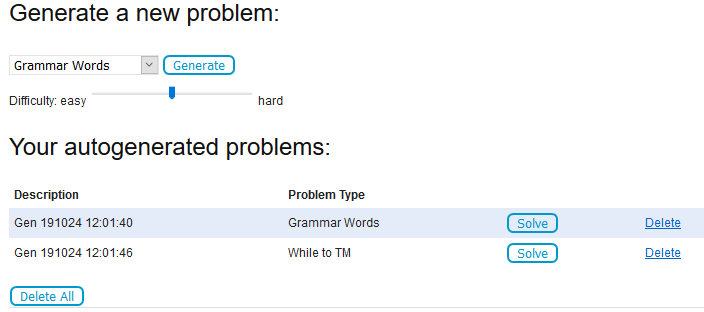}
	\caption{Automatic problem generation.}
	\label{fig:scAuto}
\end{figure}

\clearpage

\section{Complete student survey results}\label{app:survey}

The following pages show the full results of the student survey that is described in Section \ref{sec:userStudy2}.



\section*{Supplementary material}

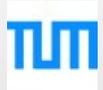

kursorganisation@prolehre.tum.de

AutomataTutor_Eval_SoSe\'19 (SoSe 2019)
Erfasste Fragebögen = 139

---

## Survey Results

## Legend

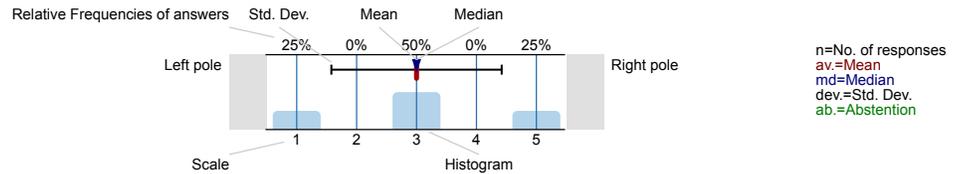

Question text

---

## Welcome to the evaluation of the Automata Tutor

Thank you for taking the time to evaluate the Automata Tutor (in the following AT) used in this course!

The evaluation will take about 7-9 minutes to fill out and will help improve the AT for future students at the TU München and other universities.

Best regards,
Maximilian Weininger
(Automata Tutor Support)

## General Usage

Did you ever use the Automata Tutor?

| | | |
|---|---|---|
| Yes. | 100% | n=138 |
| No. | 0% | |

---

( If yes, continue straight to "3. Usability of Automata Tutor (AT)" )

Why didn't you use the Automata Tutor?

| | | |
|---|---|---|
| Not applicable - I did use it. | 15.1% | n=139 |
| I didn't know of its existence. | 0% | |
| I tried and didn't like it. | 0% | |
| I found it too complicted. | 0% | |
| I didn't need it for my learning. | 0.7% | |
| For other reasons (Please provide specifics in 2.3) | 0% | |

---

## Usability of the Automata Tutor (AT)

I quickly learned to use the AT.

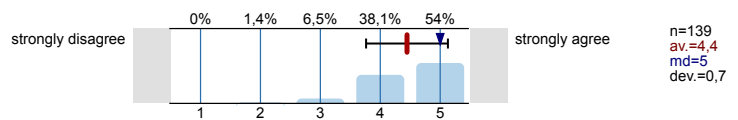

strongly disagree / strongly agree

n=139
av.=4,4
md=5
dev.=0,7

---

I need assistance to use the AT.

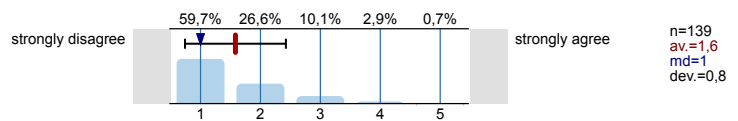

strongly disagree / strongly agree

n=139
av.=1,6
md=1
dev.=0,8

---

The AT is unnecessarily complex.

strongly disagree | strongly agree

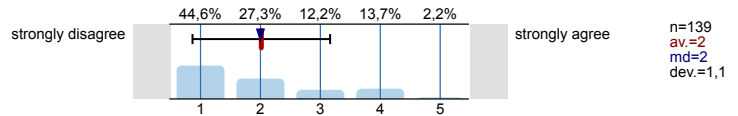

44,6% 27,3% 12,2% 13,7% 2,2%

n=139
av.=2
md=2
dev.=1,1

---

The canvas for drawing is intuitive.

strongly disagree | strongly agree

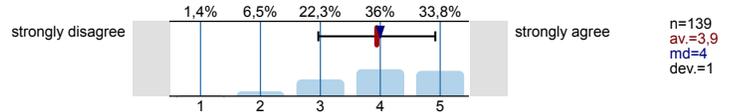

1,4% 6,5% 22,3% 36% 33,8%

n=139
av.=3,9
md=4
dev.=1

---

I feel confident using the AT.

strongly disagree | strongly agree

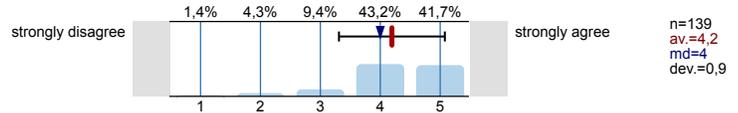

1,4% 4,3% 9,4% 43,2% 41,7%

n=139
av.=4,2
md=4
dev.=0,9

---

The AT is easy to use.

strongly disagree | strongly agree

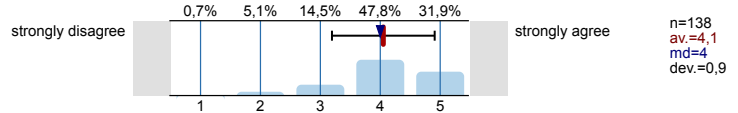

0,7% 5,1% 14,5% 47,8% 31,9%

n=138
av.=4,1
md=4
dev.=0,9

---

I enjoy using the AT/ the AT is fun to use.

strongly disagree | strongly agree

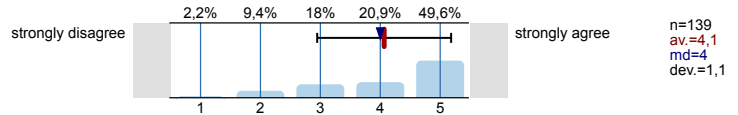

2,2% 9,4% 18% 20,9% 49,6%

n=139
av.=4,1
md=4
dev.=1,1

---

The use of AT is self-explanatory.

strongly disagree | strongly agree

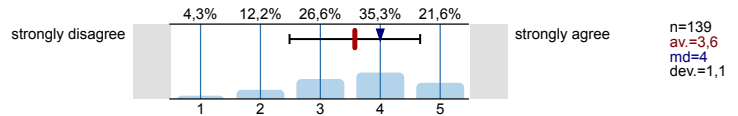

4,3% 12,2% 26,6% 35,3% 21,6%

n=139
av.=3,6
md=4
dev.=1,1

---

## Learning support of the Automata Tutor (AT)

I understand more after using the AT.

strongly disagree | strongly agree

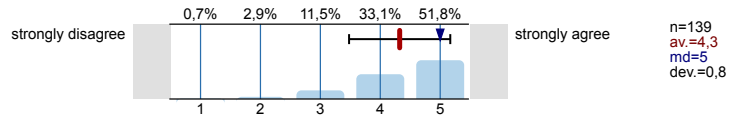

0,7% 2,9% 11,5% 33,1% 51,8%

n=139
av.=4,3
md=5
dev.=0,8

---

I prefer using the AT to using pen and paper exercises.

strongly disagree | strongly agree

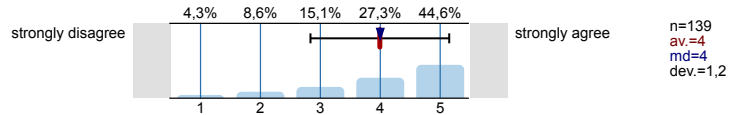

4,3% 8,6% 15,1% 27,3% 44,6%

n=139
av.=4
md=4
dev.=1,2

---

The feedback of the AT was helpful and instructive.

strongly disagree | strongly agree

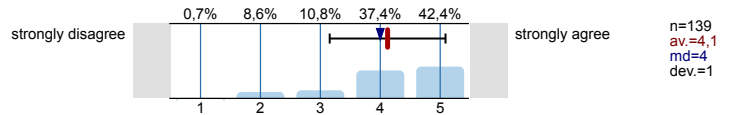

0,7% 8,6% 10,8% 37,4% 42,4%

n=139
av.=4,1
md=4
dev.=1

---

The exercises within the AT are well-designed.

strongly disagree | strongly agree

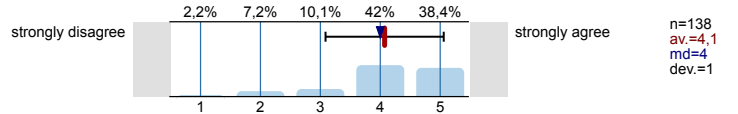

2,2% 7,2% 10,1% 42% 38,4%

n=138
av.=4,1
md=4
dev.=1

---

The AT fits in well with the programming tasks and written homework.

strongly disagree | strongly agree

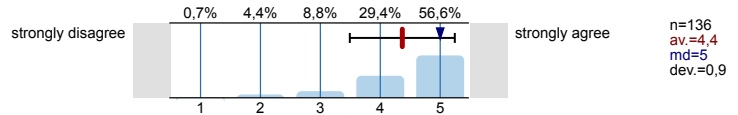

0,7% 4,4% 8,8% 29,4% 56,6%

n=136
av.=4,4
md=5
dev.=0,9

---

The AT hindered my learning.

strongly disagree | strongly agree

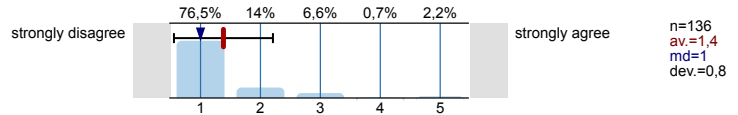

76,5% 14% 6,6% 0,7% 2,2%

n=136
av.=1,4
md=1
dev.=0,8

---

I feel better prepared for the exam after using AT.

strongly disagree | strongly agree

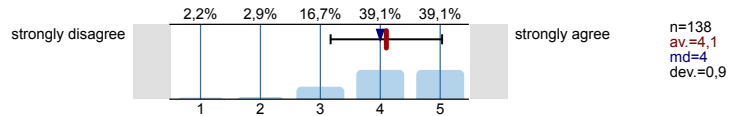

2,2% 2,9% 16,7% 39,1% 39,1%

n=138
av.=4,1
md=4
dev.=0,9

---

The feedback of the AT was misleading/ confusing.

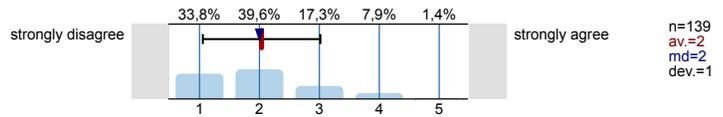

How often did you use the AT?
(Please provide only one answer).

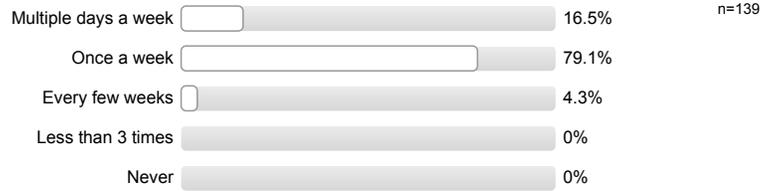

## Individual learning behaviour

How many hours a week did you invest in your homework on AT?
(Please provide only one answer).

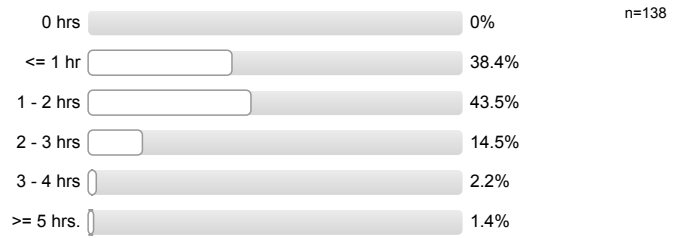

What are your preferred means of learning?
(Multiple answers possible.)

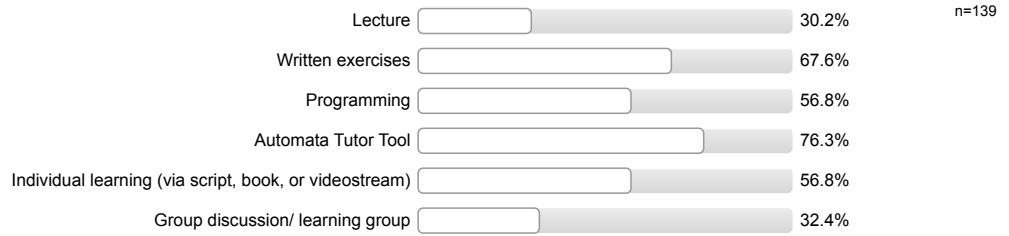

I feel well-prepared for the exam.

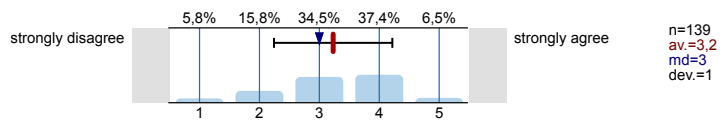

I expect a good result in my exam.

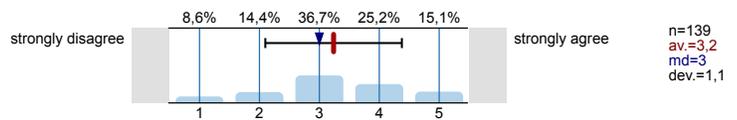

## Send data

Please do not forget to press "absenden"/"submit" so your entries will be saved and submitted.

Thank you for providing your feedback to the Automata Tutor   :-)

# Profile

## Usability of the Automata Tutor (AT)

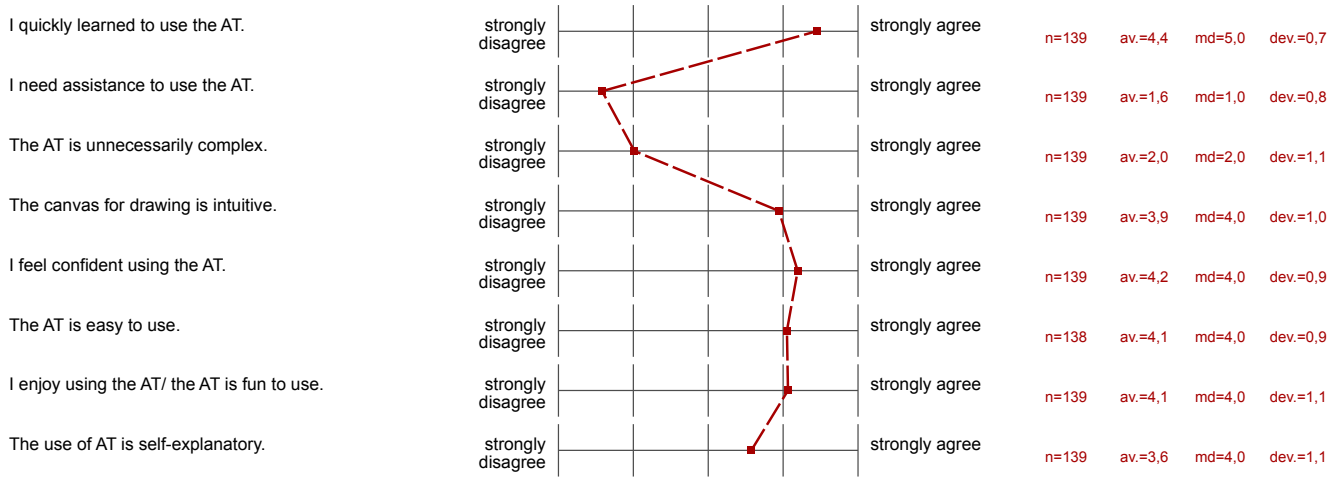

| Statement | | n | av. | md | dev. |
|---|---|---|---|---|---|
| I quickly learned to use the AT. | | n=139 | av.=4,4 | md=5,0 | dev.=0,7 |
| I need assistance to use the AT. | | n=139 | av.=1,6 | md=1,0 | dev.=0,8 |
| The AT is unnecessarily complex. | | n=139 | av.=2,0 | md=2,0 | dev.=1,1 |
| The canvas for drawing is intuitive. | | n=139 | av.=3,9 | md=4,0 | dev.=1,0 |
| I feel confident using the AT. | | n=139 | av.=4,2 | md=4,0 | dev.=0,9 |
| The AT is easy to use. | | n=138 | av.=4,1 | md=4,0 | dev.=0,9 |
| I enjoy using the AT/ the AT is fun to use. | | n=139 | av.=4,1 | md=4,0 | dev.=1,1 |
| The use of AT is self-explanatory. | | n=139 | av.=3,6 | md=4,0 | dev.=1,1 |

## Learning support of the Automata Tutor (AT)

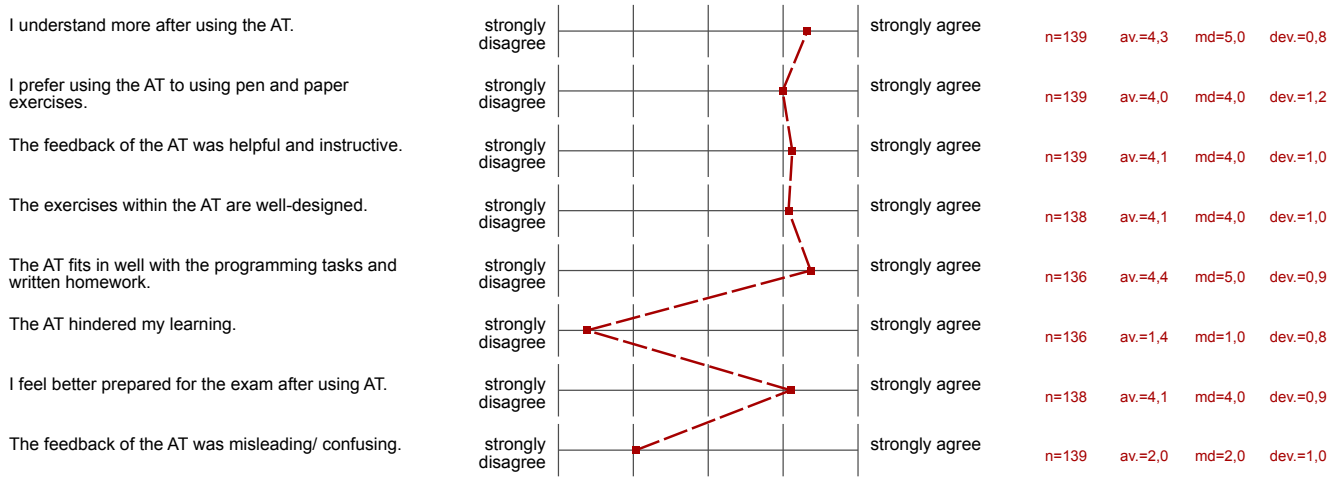

| Statement | | n | av. | md | dev. |
|---|---|---|---|---|---|
| I understand more after using the AT. | | n=139 | av.=4,3 | md=5,0 | dev.=0,8 |
| I prefer using the AT to using pen and paper exercises. | | n=139 | av.=4,0 | md=4,0 | dev.=1,2 |
| The feedback of the AT was helpful and instructive. | | n=139 | av.=4,1 | md=4,0 | dev.=1,0 |
| The exercises within the AT are well-designed. | | n=138 | av.=4,1 | md=4,0 | dev.=1,0 |
| The AT fits in well with the programming tasks and written homework. | | n=136 | av.=4,4 | md=5,0 | dev.=0,9 |
| The AT hindered my learning. | | n=136 | av.=1,4 | md=1,0 | dev.=0,8 |
| I feel better prepared for the exam after using AT. | | n=138 | av.=4,1 | md=4,0 | dev.=0,9 |
| The feedback of the AT was misleading/ confusing. | | n=139 | av.=2,0 | md=2,0 | dev.=1,0 |

## Individual learning behaviour

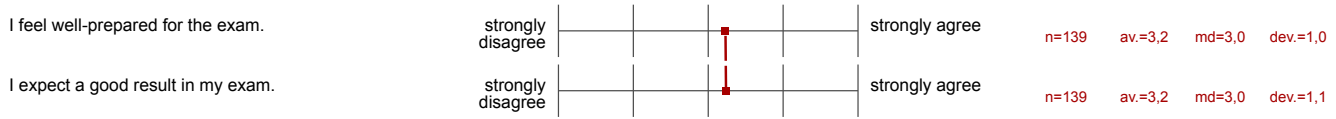

| Statement | | n | av. | md | dev. |
|---|---|---|---|---|---|
| I feel well-prepared for the exam. | | n=139 | av.=3,2 | md=3,0 | dev.=1,0 |
| I expect a good result in my exam. | | n=139 | av.=3,2 | md=3,0 | dev.=1,1 |


}
{}

\end{document}